\documentclass[conference,letterpaper]{IEEEtran}

\usepackage[utf8]{inputenc} 
\usepackage[T1]{fontenc}
\usepackage{url}
\usepackage{ifthen}
\usepackage{cite}
\usepackage[cmex10]{amsmath} 

\usepackage{amsthm}
\usepackage{amsmath}
\usepackage{enumitem}

\usepackage{amsmath,amssymb,amsfonts}
\usepackage{comment}

\usepackage{textcomp}
\usepackage[table,xcdraw]{xcolor}
\usepackage{tikz}

\usepackage{graphicx}
\usepackage{subcaption}

\newtheorem{theorem}{Theorem}

\newtheorem{remark}{Remark}
\newtheorem{definition}{Definition}

\newtheorem{example}{Example}

\newcommand{\red}[1]{\textcolor{red}{#1}}

\newcommand{\bF}{\mathbb{F}}

\newcommand{\cA}{\mathcal{A}}

\newcommand{\cI}{\mathcal{I}}

\newcommand{\cL}{\mathcal{L}}

\newcommand{\cR}{\mathcal{R}}

\newcommand{\cT}{\mathcal{T}}

\newcommand{\boldx}{\mathbf{x}}

\newcommand{\boldz}{\mathbf{z}}

\newcommand{\off}[1]{}
\DeclareMathOperator*{\argmax}{arg\,max} 

\begin{document}

\title{Individual Confidential Computing of Polynomials over Non-Uniform Information\vspace{-0.4cm}} 


\author{%
   \IEEEauthorblockN{\textbf{Saar Tarnopolsky}\IEEEauthorrefmark{1},
                    \textbf{Zirui (Ken) Deng}\IEEEauthorrefmark{2},
                    \textbf{Vinayak Ramkumar}\IEEEauthorrefmark{3},
                    \textbf{Netanel Raviv}\IEEEauthorrefmark{2},
                     and \textbf{Alejandro Cohen}\IEEEauthorrefmark{1}}
   \IEEEauthorblockA{\IEEEauthorrefmark{1}%
                      Faculty of ECE, Technion, Israel, \texttt{saar@campus.technion.ac.il},  \texttt{alecohen@technion.ac.il}}  
    \IEEEauthorblockA{\IEEEauthorrefmark{2}%
                     Department of Computer Science and Engineering, Washington University in St. Louis, St. Louis, USA,\\
                     \texttt{d.ken,netanel.raviv@wustl.edu}\\
    \IEEEauthorrefmark{3}Department of Electrical Engineering--Systems, Tel Aviv University, Israel, \texttt{vinram93@gmail.com}}
\vspace{-1.1cm}
}

\maketitle


\begin{abstract}

In this paper, we address the problem of secure distributed computation in scenarios where user data is not uniformly distributed, extending existing frameworks that assume uniformity, an assumption that is challenging to enforce in data for computation. Motivated by the pervasive reliance on single service providers for data storage and computation, we propose a privacy-preserving scheme that achieves information-theoretic security guarantees for computing polynomials over non-uniform data distributions. Our framework builds upon the concept of perfect subset privacy and employs linear hashing techniques to transform non-uniform data into approximately uniform distributions, enabling robust and secure computation. We derive leakage bounds and demonstrate that information leakage of any subset of user data to untrusted service providers, i.e., not only to colluding workers but also (and more importantly) to the admin, remains negligible under the proposed scheme. 
\end{abstract}

\section{Introduction}
\off{The widespread adoption of third-party services for data storage and computation introduces significant privacy and security risks. Coded computing methods have been developed in order to tackle privacy issues in these systems~\cite{lee2017speeding,yu2019lagrange,d2020gasp,wang2022breaking}, but they typically provide privacy guarantees in terms of some level of restricted collusion among workers, while the system administrator is assumed to be trusted.}

The widespread adoption of third-party services for data storage and computation poses significant privacy and security risks. Various approaches were considered for this crucial problem\off{, including Differential Privacy, Privacy-Utility Tradeoffs, Perfect Privacy, and Confidential Computing}. For example, Differential Privacy measures a service provider's ability to infer user data by analyzing changes in output statistics caused by altering a single data point \cite{cuff2016differential,mironov2009computational,huang2019dp}. Privacy-Utility Tradeoffs involve users providing distorted data to the service provider, with security measured by the level of service achievable given the distortion\off{; naturally, greater distortion limits utility} \cite{li2009tradeoff,wang2017privacy}. Perfect Privacy considers user-encoded data and evaluates security based on the mutual information between the confidential and encoded data \cite{calmon2015fundamental,rassouli2017perfect,rassouli2021perfect}. In computer science, similar security risks are addressed under the framework of Confidential Computing~\cite{mulligan2021confidential,hunt2021confidential,feng2024survey}. This approach ensures the isolation of confidential data across untrusted service providers through hardware and software guarantees.

\begin{figure}
    \centering
    \includegraphics[width=1\linewidth]{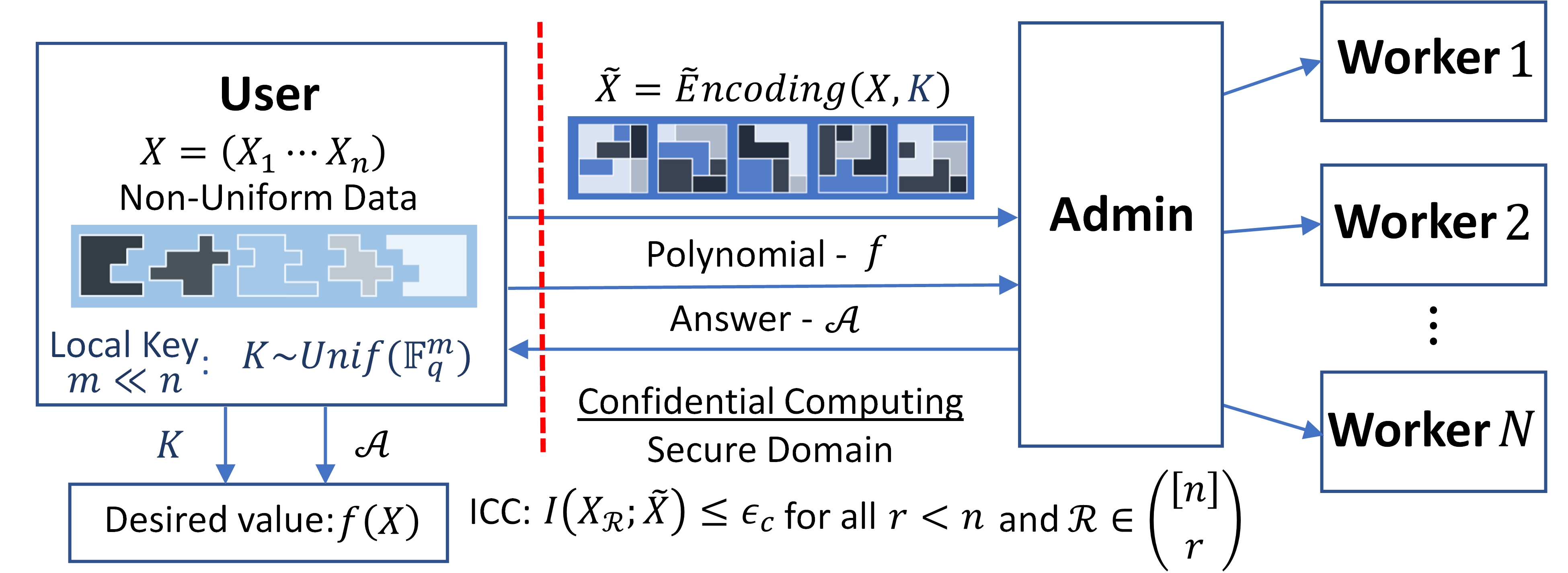}
    \caption{\small Individual confidential computing (ICC) scheme for non-uniform data, $X\in \bF_q^n$, drawn from some unknown distribution. Red dashed line represents the ``security barrier'' in ICC scheme. Instead of protecting the admin's data against colluding workers \cite{lee2017speeding,wang2022breaking,yu2019lagrange,d2020gasp}, ICC protects the user's data against the service provider as a whole.}
    \label{fig:icc_scheme}
    \vspace{-0.7cm}
\end{figure}

In this paper, we consider the framework of Perfect Privacy, or more specifically Perfect Subset Privacy~\cite{raviv2022perfect}. Coded computing methods have been developed to tackle privacy issues in these framework~\cite{lee2017speeding,yu2019lagrange,d2020gasp,wang2022breaking}, but they typically provide privacy guarantees in terms of some level of restricted collusion among workers, while the system administrator is assumed to be trusted. Recent work~\cite{deng2024perfect} highlights a frequently overlooked yet common scenario in which a data owner entrusts their data to a single \textit{untrusted} service provider for storage and computation. In this setting, illustrated in Fig.~\ref{fig:icc_scheme}, the service provider presents an inherent privacy risk by having full access to the data. This remains true regardless of whether storage and computation are internally distributed, effectively nullifying any assumptions about limited collusion among workers. This single-provider scenario represents a more realistic and prevalent setting for modern cloud computing services, necessitating novel approaches to secure computation.

Ref.~\cite{deng2024perfect} broadens the scope of coded computing by developing techniques that protect privacy against the entire service provider. Specifically, \cite{deng2024perfect} considers a setting where the user possesses uniformly distributed data. The service provider comprises a system administrator and multiple workers, with no restrictions on potential collusion among workers. The user cannot communicate directly with the workers and must interact solely through the system administrator. To safeguard their data, the user encodes the data using a random key and sends the encoded version to the administrator, who then distributes it to the workers. Due to storage constraints, the user retains only the key as side information and relies on information exchange with the system administrator to retrieve the desired computation results, potentially utilizing the random key in the process. 

In this work, we generalize~\cite{deng2024perfect} by considering user data that is \textit{not necessarily uniformly} distributed and provide information-theoretic security guarantees accordingly. The privacy framework introduced in~\cite{deng2024perfect} employs the recently developed concept of perfect subset privacy~\cite{raviv2022perfect}, which ensures zero leakage from any subset of the data up to a specified size. However, when dealing with non-uniform user data, achieving zero leakage becomes impossible without local randomness at the user's data entropy (see Section~\ref{section:securitymetrics}). Therefore, in this paper, we aim to guarantee limited information leakage instead of absolute zero leakage, which, as demonstrated in \cite{10619109}, \emph{meets the cryptographic criterion for ineligibility} \cite{bellare1998relations,enwiki:indistinguishability,enwiki:Negligible}. 
To achieve this, we utilize the smoothing technique introduced by M. Pathegama et. al. \cite{pathegama2024r}, which enables the transformation of an unknown source with a given entropy level into a uniform random variable using a random linear code and a uniform key. This transformation enables the application of coded computing methods for secure distributed computation in non-uniform settings.

We introduce the concept of \textit{Individual Confidential Computing (ICC)} to extend secure computation frameworks further. Unlike traditional privacy settings where adversaries observe only a subset of the encoded data, ICC assumes that the entire encoded message is accessible to untrusted service provider. ICC ensures that no information about any subset of the user’s data, up to a specified size, leaks to the service provider. This definition broadens the applicability of secure computation techniques, as it accommodates the realistic assumption of full data access by the service provider.

The remainder of the paper is structured as follows. The general computation model for this work is given in Section~\ref{section:model}. The privacy and security metrics under consideration are discussed in Section~\ref{section:securitymetrics}. Preliminary results from previous works are provided in Section~\ref{section:prelim}. The main results of this work are presented in Section~\ref{section:mainresults}.

\section{Computation System Model}\label{section:model}
Consider a user possessing data~$X=(X_1, \ldots, X_n) \in \bF_q^n$, drawn from some unknown distribution. As illustrated in Fig.~\ref{fig:icc_scheme}, the user wishes to share data with a third-party service provider in order to compute a polynomial~$f$ on~$X$ at a later point in time, and~$f$ is not known prior to sharing of data.\off{
\red{[NR: Add ``The user wishes to share~$X$ with a third party service provider in order to compute a polynomial~$f$ on~$X$ at later point in time, where~$f$ is not known prior to sharing~$X$.'']}}
The user encodes~$X$ into~$\tilde{X} = \tilde{E}(X,K)$, where $K\sim \textrm{Unif}(\bF_{q}^{m})$ is a randomly generated secret key with~$m \ll n$.  The encoded data~$\tilde{X}$ is then transmitted to the system administrator of the service provider, who subsequently distributes it to the workers within the system. The user intends to discard the original data~$X$ and retain only the secret key~$K$ as side information, thereby minimizing storage requirements. This process is referred to as the \textit{storage phase} of the model.

At a later point, the user is interested in computation of some specific polynomial~$f$ over their original data, and they share the identity of~$f$ with the system admin. The polynomial~$f$ is not known during the storage phase. Upon receiving~$f$, the admin coordinates the workers to perform 
some computation over their share of the encoded data and return the results. The administrator then aggregates these results and sends them to the user. Utilizing this aggregated information in conjunction with the secret key~$K$, the user performs the final decoding to retrieve the desired value~$f(X)$. We call this the \textit{computation phase} of the model.
Among all workers, there may exist some straggling ones that fail to respond in a timely manner, and the admin must ensure that sufficient information is available for decoding~$f(X)$ even in the presence of such stragglers. 

The primary distinction in our work lies in the handling of non-uniformly distributed data~$X$. Unlike in~\cite{deng2024perfect}, where $X$ is assumed to be uniformly distributed, in this work our approach guarantees secure polynomial computation over non-uniform user data, thereby extending the applicability and robustness of coded computing techniques.

We say that any scheme that realizes this computation model is an~\textit{$(n, q, r, d, S)$ scheme}, where~$n$ is the length of the data held by the user, $q$ is the field size, $r < n$ is the security parameter (see Section~\ref{section:securitymetrics} for details)\off{ \red{[NR: privacy metric not introduced yet. Either introduce it here or promise the reader that the meaning of~$r$ will be explained in Section~XX]}}, $d$ is the maximum total degree of~$f$, and~$S$ is the maximum number of stragglers in the system. We judge the merit of such schemes by three quantities: the number of workers needed, measured by the parameter~$N$; download cost, measured by the number~$D$ of~$\bF_q$ symbols the user needs to download in order to complete the decoding process; side information, measured by~$m$, the size of the random key~$K$.

\section{Privacy and Security Metrics}\label{section:securitymetrics}

In his seminal 1949 work~\cite{shannon1949communication}, Shannon introduced the concept of perfect secrecy for communication between two legitimate users, Alice and Bob, in the presence of an eavesdropper, Eve. Shannon defined information-theoretic \textit{perfect secrecy} as an encoding that satisfies $I(\tilde{X};X) = 0$, where $X$ is the secret message sent from Alice to Bob, $\tilde{X}$ is the encoding of $X$ observed by Eve, and $I(\cdot;\cdot)$ denotes the mutual information function. In other words, the encoded message is independent of the original message. Shannon demonstrated that this level of secrecy can only be achieved if Alice and Bob share a secret key with entropy at least as large as the entropy of the message, i.e., $H(K) \geq H(X)$.

Sharing large secret keys per transmission between Alice and Bob is not practical in real communication systems. Consequently, relaxations of perfect secrecy have been proposed to enable more practical secrecy measures. One such relaxation involves limiting Eve's observations of the encoded message. This approach, known as \textit{physical layer security} \cite{bloch2011physical}, exploits Eve's weaker observations to achieve security at the expense of the communication rate. An example of such a scheme is the Wiretap Channel of Type II, introduced by Ozarow and Wyner in~\cite{ozarow1985wire}. In their scheme, Eve observes $w$ out of $n$ transmitted symbols of the encoded message, denoted by $\tilde{X}_w$. Ozarow and Wyner demonstrated that, by using a local source of randomness, Eve cannot gain any information about the encoded message from her observations, i.e., $I(\tilde{X}_w;X) = 0$, provided that Alice transmits at a rate lower than $\frac{n-w}{n}$.

To increase the secure communication rate, another relaxation was proposed, referred to as \textit{individual secrecy} (IS)~\cite{carleial1977note,bhattad2005weakly}. In this setting, Alice aims to transmit $\ell$ messages to Bob: $X_1, \dots, X_{\ell}$, where Eve can observe any $w<\ell$ encoded messages denoted by $\underline{\tilde{X}}_w$. Individual secrecy guarantees that $\{I(\underline{\tilde{X}}_w; X_i)=0\}_{i=1}^{\ell}$. That is, Eve cannot obtain information about any individual message but may gain an insignificant amount of information about the combination of messages $X_1, \dots, X_{\ell}$ \cite{cohen2018secure}. Although a negligible amount of information may leak about the combination of messages, Eve has no knowledge of any individual message, while Alice can transmit at the full communication rate. IS can also be extended to sets of messages, i.e., $I(\underline{\tilde{X}}_w; \underline{X}_{\ell-w}) = 0$ for any subset of $\ell-w$ messages from $X_1, \dots, X_{\ell}$. Unlike previous notions, IS does not require Alice to use a secret random key for encoding. Instead, Alice mixes the messages, using a subset of them to protect the remaining ones. However, this approach implies that, unlike earlier schemes, Alice must compress the messages prior to encoding in order to handle non-uniform information and achieve individual secrecy~\cite{matsumoto2017universal,10619109}.

In this paper, we aim to provide secrecy in computation \cite{sevilla2022compute,egger2022efficient,xhemrishi2022computational}. In the literature, this concept is often referred to as Perfect Privacy~\cite{calmon2015fundamental,rassouli2017perfect}, or Confidential Computing in the computer science society~\cite{mulligan2021confidential,hunt2021confidential,feng2024survey}. In these settings, a legitimate user seeks to perform some computation on a random variable $X$, and the computation is delegated to an \textit{untrusted} service provider which consists of an admin and workers (see Fig.~\ref{fig:icc_scheme}). The user encodes $X$ into $\tilde{X}$ s.t. the admin and workers cannot obtain any information about $X$ from $\tilde{X}$, i.e., $I(\tilde{X};X) = 0$. This definition resembles perfect secrecy and similarly requires encoding $X$ using a large secret key. Perfect Confidential Computing is a very strong notion that requires a key as big as the data itself. Thus, relaxations for this setting have also been considered, e.g., Privacy-Utility tradeoffs, Differential Privacy, and more \cite{dwork2006differential,li2009tradeoff,wang2017privacy,rassouli2019data}. 

In this work, we focus on \textit{Individual Confidential Computing (ICC)}. In ICC, the user performs computation on a random variable $X \in \mathbb{F}_{q}^{n}$ using an untrusted admin and workers. However, ICC ensures that the untrusted admin and workers cannot obtain any information about any subset of size $1 \leq r < n$ symbols of $X$ from observing the entire encoded message $\tilde{X}$, where~$r$ is a tunable security parameter. Below, we provide the formal definition of ICC.

\begin{definition}[Individual Confidential Computing]\label{definition:ICC}
    Let $X \in \mathbb{F}_{q}^{n}$ be a random variable with distribution $p_X$, $1 \le r < n$ and $[n] \triangleq \{1,\ldots,n\}$ be a security parameter. An encoding scheme is said to be Individually Confidential Computing (ICC) if $I(X_{\mathcal{R}};\tilde{X}) \leq \epsilon_c$ for all $\cR \in \binom{\left[n\right]}{r}$ and some negligible~$\epsilon_c>0$.
\end{definition}

\begin{remark}    
The definition of ICC is equivalent to the concept of $r$-subset privacy considered in~\cite{raviv2022perfect,deng2024perfect} for uniform information. Moreover, an important distinction between ICC definition and IS is that, in IS, Eve may observe only a subset of the encoded information, whereas in ICC, the untrusted admin and workers have access to the entire encoded message. To achieve this level of security, even with uniform information, ICC schemes must utilize a random key of at least size $r$ as demonstrated in~\cite{raviv2022perfect,deng2024perfect}.
\end{remark}

\begin{remark} \label{rm:epsilon_and_m}
    The key size~$m$  for an \textit{$(n, q, r, d, S)$} computation model is a function of the selected~$\epsilon_c$ in Definition~\ref{definition:ICC}, as shown in Theorem~\ref{theorem:main_theorem} and illustrated in Example~\ref{exam:key_size} (see Figs.~\ref{fig:graph-leakge-key} and~\ref{fig:graph-entropy-key}). 
\end{remark}

\section{Preliminaries}\label{section:prelim}
\subsection{Smoothing}
\label{sec:smoothing}
In this paper, we utilize smoothing of distributions \cite{micciancio2007worst,debris2023smoothing} as a method to achieve ICC (Definition~\ref{definition:ICC}) for non-uniform information. Smoothing of distributions is used a technique to transform a non-uniform random variable $X \in \mathbb{F}_{q}^{n}$ into an almost uniform random variable. Due to space limitation we provide here a brief overview of the connection between smoothing and ICC (Definition~\ref{definition:ICC}), and refer to \cite[Appendix A]{ICC2025} for a more detailed description.

We start by providing a formal definition of smoothing of distributions as also used in \cite{pathegama2024r}. We use the $p$-variational distance denoted by $\mathbb{V}_{p}(\cdot,\cdot)$ for $p \geq 1$, $p\in \mathbb{N}$ to measure the distance of the smoothed distribution to a uniform distribution.

\begin{definition}\label{def:smoothing}
    (Smoothing using Linear Codes) Let $X \in \mathbb{F}_{q}^{n}$ be some non-uniform random variable with distribution $X \sim p_X$. Let $C = [n,m]_{q}$ be a linear code. Let $\mathcal{C} \in \mathbb{F}_{q}^{n}$ be a uniformly randomly chosen codeword from~$C$, and $\tilde{X} = X + \mathcal{C}$. Denote the distribution $\tilde{X} \sim p_{\tilde{X}}$. We say the code $C$ is smoothing the distribution $X \sim p_{X}$ if
        $\mathbb{V}_{p}\left(p_{\tilde{X}},p_{U_n}\right) \leq \epsilon_s$
    for some negligible $\epsilon_s > 0$, where $p_{U_n}$ is the uniform distribution over $\mathbb{F}_{q}^{n}$. 
\end{definition}

Thus, the encoding $\tilde{X} = X + \mathcal{C}$, where $\mathcal{C}$ is chosen uniformly from $C$, generates a random variable $\tilde{X}$ that is nearly uniformly distributed over $\mathbb{F}_{q}^{n}$.

In Thm. 3.1 of \cite{pathegama2024r}, M. Pathegama et. al., provide the condition under which a random linear code $C = [n, m]_q$ with a generator matrix $G \in \mathbb{F}_{q}^{m \times n}$ and $K \sim \text{Unif}(\mathbb{F}_q^m)$ performs smoothing on a non-uniform distribution $X \sim p_{X}$, where $X \in \mathbb{F}_q^n$, through the encoding $\tilde{X} = X + KG$. Specifically, the theorem asserts that for $p \geq 2$, and negligible $\epsilon > 0$, taking  $m$ larger than a function of $n, p, \log_q(1/\epsilon)$, and $H_p(X)$, ensures that with high probability, a randomly chosen linear code $C = [n, m]_q$ performs smoothing of $X \sim p_{X}$ s.t.
\vspace{-0.1cm}
\begin{gather*}
    \mathbb{V}_{p}(p_{\tilde{X}},p_{U_n}) \leq 2^{\frac{p-1}{p}}((1+\epsilon)^{p}-1)^{\frac{1}{p}}.
\end{gather*}

In this paper, we apply this result within the context of ICC. ICC quantifies information leakage through mutual information. Thus, for negligible information leakage, ICC requires the encoded information and the distribution of some subset of the data to be nearly independent. This requirement is equivalent to asking for the distributions $p_{\tilde{X}}$ and $p_{\tilde{X}|X_{\mathcal{R}}}$ to be close to each other, when $\mathcal{R}$ is some subset of size~$r$ of the encoded data. In Sec.~\ref{sec:VD_theorem} and~\ref{sec:MI_theorem}, smoothing and \cite[Thm.~3.1]{pathegama2024r} are used to ensure the required distributions are nearly uniform, which results in negligible leakage and consequently the satisfaction of ICC conditions.

\subsection{Code-Based Polynomial Computation Scheme} \label{sec:coded_based_linear_comp}

For completeness, we summarize the $(n, q, r, d, S)$ scheme used in~\cite{deng2024perfect} based on information super-sets for Reed-Muller codes. The scheme improves upon a trivial decoding scheme that is equivalent to trying all possible guesses for the correct random key. 

An \textit{information set}~$\cI$ for a code~$C$ is a subset of positions within any codeword of~$C$ that uniquely determines the entire codeword. A multiset~$\cT$ is called an \textit{$S$-information super-set} for a code $C$ if every~$(|\cT|-S)$-subset of~$\cT$ contains an information set for~$C$. For parameters~$d$ and~$m$, denote $RM_q(d, m) = \{\textrm{Eval}(h)\mid h \in \bF_q[\boldx], \deg(h) \le d\}$ as the Reed-Muller code over $\bF_q$, where $\deg(\cdot)$ denotes the total degree of a polynomial, and~$\textrm{Eval}(h)=(h(\boldz))_{\boldz\in \bF_q^m}$.

Provided the total degree~$d$ of the polynomial~$f$ satisfies~$d<m(q-1)$, where~$m$ is the dimension of the linear code used in the current paper for smoothing,
the information super-set based $(n, q, r, d, S)$-scheme in~\cite{deng2024perfect} can be described as follows.
Let~$\cT$ be an $S$-information super-set for $RM_q(d,m)$ with size~$|\cT| = N$. Let $G \in \mathbb{F}_{q}^{m \times n}$ be the generator matrix of the linear code $C$. 
\begin{enumerate}[label=\roman*)]
    \item The user encodes~$X$ to $\Tilde{X} = X + K G$ for $K \sim \textrm{Unif}(\bF_{q}^m)$, and sends $\Tilde{X}$ to the admin. 
    \item The admin encodes~$\Tilde{X}$ to  ${\hat{X}} = (\hat{X}_1, \ldots, \hat{X}_N) = ({\Tilde{X}} - T G | T \in \cT)$ 
    and distributes them to a total of $N$ workers.
    \item At a later point, the user sends $f$ to the admin, who then sends it to the workers. The workers apply~$f$ on their data, and send the results back to the admin.
    \item The admin obtains $\{f({\Tilde{X}} - T G)\}_{T \in \cI}$, where~$\cI \subseteq\cT$ is an information set for~$RM_q(d,m)$,
    and sends $\cA \triangleq (f({\Tilde{X}} - T G))_{T \in \cI}$ to the user. 
    \item The user linearly combines the values in~$\cA$ to obtain~$f(X)$.
\end{enumerate} 
To see why this scheme works, define multivariate polynomial~$g: \bF_q^m \rightarrow \bF_q$ as $g(T) \triangleq f({\Tilde{X}} - T G)$.
From the definition of information super-sets, it follows that even in the presence of~$S$ stragglers, the admin still receives the evaluations of~$g$ at an information set~$\cI$ for~$RM_q(d,m)$. As a result, the admin can compute the answer $\cA$ by evaluating~$g$ at~$\cI$.

This information super-set based scheme has download cost~$D$ equal to the dimension of~$RM_q(d,m)$, an improvement over the trivial scheme that requires a download cost of~$D = q^m$. In terms of number of workers needed, the above scheme using information super-sets needs~$N = L(q,d,m,S)$ workers, where~$L(q,d,m,S)$ is the minimum size of an $S$-information super-set for $RM_q(d,m)$, while the trivial scheme requires~$N = (S+1)D$ workers. 
Refer to~\cite{deng2024perfect} for details about constructions and bounds on information super-sets for Reed-Muller codes, which further improve over the trivial scheme.
\vspace{-0.1cm}
\begin{figure}
     \centering
     \begin{subfigure}[b]{0.49\textwidth}
         \centering
         \includegraphics[width=0.95\textwidth]{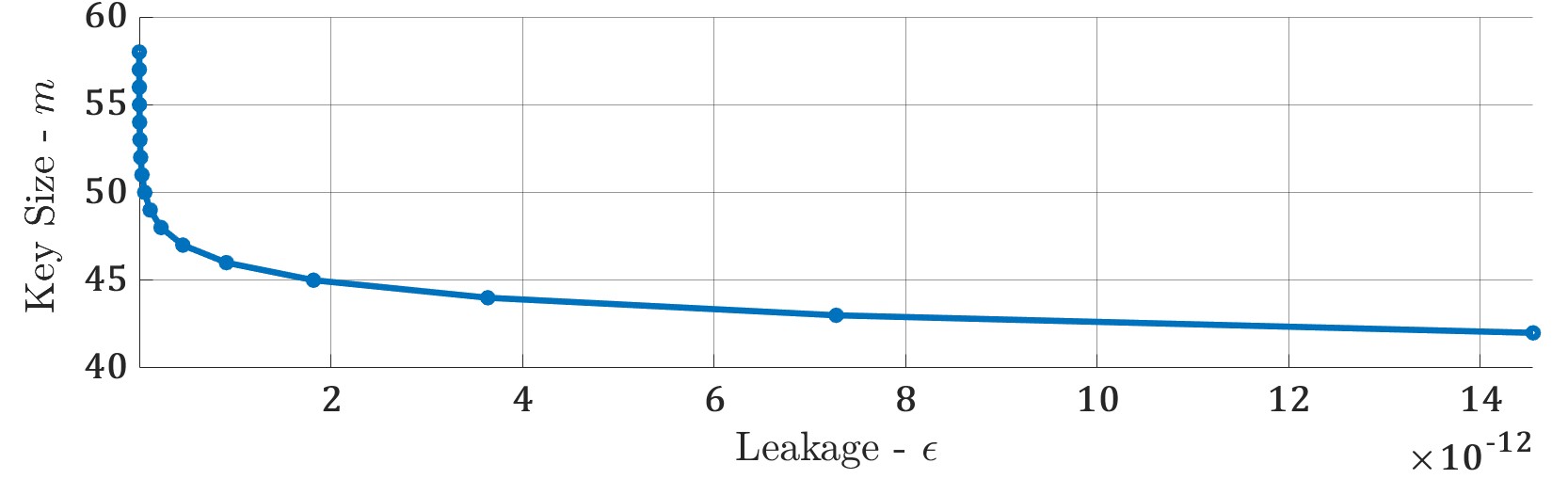}
         \caption{$m$ as function of information leakage $\epsilon$ for data entropy $n-4$, i.e. $\min_{\mathcal{R},z \in \mathbb{F}_{q}^{r}}\left(H_{p}(X|X_{\mathcal{R}}=z)\right) = n - 4$.}
         \label{fig:graph-leakge-key}
     \end{subfigure}
        \hfill
     \begin{subfigure}[b]{0.49\textwidth}
         \centering
         \includegraphics[width=\textwidth]{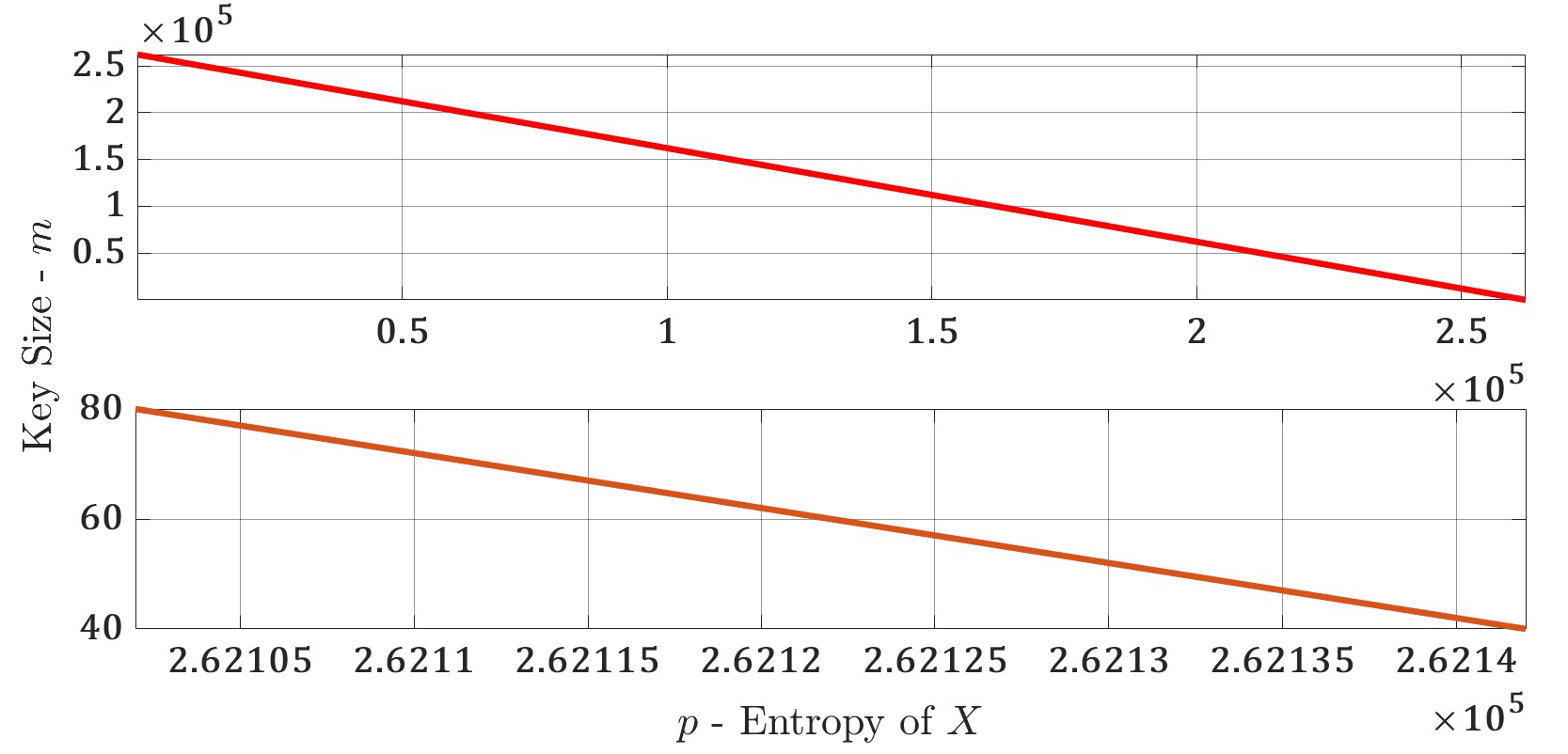}
         \caption{$m$ as function of data entropy $\min_{\mathcal{R},z\in \mathbb{F}_{q}^{r}}\left(H_{p}(X|X_{\mathcal{R}}=z)\right)$ for $\epsilon = n^{-2} \approx 10^{-12}$.}
         \label{fig:graph-entropy-key}
     \end{subfigure}
        \hfill
        \label{fig:graphs}
        \vspace{-0.5cm}
        \caption{\small Key size $m$ as a function of (a) information leakage and (b) entropy of the data, for $n = 262144 = 2^{18}$, $p=2$, $q=2$, $r=2$.}
        \vspace{-0.5cm}
\end{figure}
\section{Main Results}\label{section:mainresults}
In this section, we provide a leakage and achievability theorem for ICC (see Definition~\ref{definition:ICC}) code-based polynomial computation. We consider the linear encoding of non-uniformly distributed data for a distributed computation of a polynomial using information super-sets for Reed-Muller codes as given in Sec.~\ref{sec:coded_based_linear_comp}. 
Specifically, we bound the information leakage from any subset of size $1 \leq r < n$ of the non-uniform data $X$ to the untrusted admin and workers observing the encoded data $\tilde{X}$. We show this information leakage is negligible while the user can reliably decode the computation performed by the workers. 

\begin{theorem}\label{theorem:main_theorem}
Let $X \in \mathbb{F}_{q}^{n}$ be a non-uniform random variable with distribution $p_X$, let $a > 1$, let~$p\ge 2$ be an integer, and let~$\epsilon>0$. 
Consider a scheme in which~$X$ is encoded using an $(n,q,r,d,S)$ Code-Based Polynomial Computation scheme as in Sec.~\ref{sec:coded_based_linear_comp}, with key of size~$m$, and a linear code~$C=[n,m]_q$ with a generator matrix $G \in \mathbb{F}_{q}^{m \times n}$ chosen uniformly as random.
If
\vspace{-0.2cm}
\[
m \geq n + p + \log_{q}\left(1/\epsilon\right) - H_{p}(X) + \max_{\mathcal{R}}H_{p}(X_{\cR}),
\vspace{-0.2cm}
\]
    then with probability at least $1 - \frac{1}{a}$, the encoding $\tilde{X} = X + KG$ satisfies ICC, i.e., $I(\tilde{X};X_{\mathcal{R}}) \leq \epsilon_c$ for every~$\cR\in\binom{[n]}{r}$, with 
 \off{   \begin{align*}
        &I(\tilde{X};X_{\mathcal{R}}) \leq\\ 
        &\frac{p}{p-1}\log_{q}\left(1 + a\cdot  2^{\frac{2p-1}{p}}\left(1+q^{-\max_{\cR}H_{p}(X_\cR)}\right)\epsilon^{\frac{1}{p}}\right),
    \end{align*}}
    \begin{align*}
        \epsilon_c = \frac{p}{p-1}\log_{q}\left(1 + a\cdot  2^{\frac{2p-1}{p}}\left(1+q^{-\max_{\cR}H_{p}(X_\cR)}\right)\epsilon^{\frac{1}{p}}\right),
    \end{align*}
    and the computation results can be reliably decoded by the user.
    \end{theorem}

The leakage proofs of Theorem~\ref{theorem:main_theorem} under variational distance and mutual information are provided in Sec.~\ref{sec:VD_theorem} and Sec.~\ref{sec:MI_theorem}, respectively, whereas the reliability of the scheme is a direct consequence of the proof given in \cite{deng2024perfect}. The theorem states that when performing computation over a non-uniform random variable, ICC can be guaranteed with high probability using a randomly chosen linear code, provided that the key size is sufficiently large. The key size depends on the entropy of $X$, the amount of information leakage, the field size $q$, and the security parameter $r$. 

\begin{example}\label{exam:key_size}
Let $\epsilon = n^{-b}$ and~$r > 1$, and suppose that $\max_{\cR}(H_{p}(X_{\cR})) = r - 1$ and $H_{p}(X) = n - 1$. Then, the key size should be at least
    $m \geq r + p + b\log_{q}(n)$.
    As $n$ grows, $b\log_{q}(n)$ becomes negligible compared to $n$. Additionally, for a constant $r/n$ ratio, $b\log_{q}(n)$ becomes negligible compared to $r$ as $n$ grows. We refer to Figs.~\ref{fig:graph-leakge-key} and~\ref{fig:graph-entropy-key} to illustrate the growth of the required key size, $m$, as a function of the  information leakage $\epsilon$ (Fig.~\ref{fig:graph-leakge-key}), and the $p$-entropy of the data $\min_{\cR, z \in \mathbb{F}_{q}^{r}}\left(H_{p}(X|X_{\cR}=z)\right)$ (Fig.~\ref{fig:graph-entropy-key}).
\end{example}
\vspace{-0.1cm}
\vspace{-0.2cm}
\subsection{Leakage Proof of Theorem~\ref{theorem:main_theorem} Under Variational Distance} \label{sec:VD_theorem}
Let $X \in \mathbb{F}_{q}^{n}$ be a random variable drawn from a non-uniform distribution $p_{X}$ over $\mathbb{F}_{q}^{n}$, i.e., $X \sim p_X$, 
and let $Y \sim \text{Unif}(\mathbb{F}_{q}^{n})$.
We encode $X$ using a random key $K \sim \text{Unif}(\mathbb{F}_{q}^{m})$ and a uniformly random $[n, m]_q$ linear code, defined by the generator matrix $G \in \mathbb{F}_{q}^{m \times n}$, i.e.,
    $\tilde{X} = X + K\cdot G$. 

We start by bounding $\mathbb{V}(p_{\tilde{X}|X_{\mathcal{R}}=z}, p_{\tilde{X}})$ for any~$\mathcal{R} \in \binom{[n]}{r}$ and any~$z \in \mathbb{F}_{q}^r$, where $\mathbb{V}(\cdot, \cdot)$ denotes variational distance. For a fixed $\mathcal{R} \in \binom{[n]}{r}$ and a fixed $z \in \mathbb{F}_{q}^{r}$, the triangle inequality implies that
\vspace{-0.1cm}
\begin{align}
&\mathbb{V}\left(p_{\tilde{X}|X_{\mathcal{R}}=z},p_{\tilde{X}}\right) \leq  \mathbb{V}\left(p_{\tilde{X}|X_{\mathcal{R}}=z},p_Y\right) + \mathbb{V}\left(p_{\tilde{X}},p_{Y}\right).\label{eq:k_s_ind}
\end{align}
Therefore, to bound the variational distance, we require $\tilde{X}$ and $\tilde{X} | X_{\cR} = z$ to be close to being uniformly distributed.
Thus, we require the generator matrix~$G$ to perform smoothing of the distributions~$X$ and $X|X_{\cR} = z$ as defined in Definition~\ref{def:smoothing}. 

Let $\epsilon > 0$ (see Remark~\ref{rm:epsilon_and_m}) 
be some negligible number, and let $p \geq 2$ be some positive integer. 
According to~\cite[Thm.~3.1, Lem.~3.6]{pathegama2024r}, \off{there exists a generator matrix $G \in \mathbb{F}_{q}^{m \times n}$ such that} if 
\vspace{-0.1cm}
\begin{gather} \label{eq:key_size}
    m \geq n + p + \log_{q}\left(\frac{1}{\epsilon}\right) - H_{p}(X|X_{\cR} = z),
\end{gather}
then the expectation of the $p$-variational distance ($\mathbb{V}_{p}(\cdot,\cdot)$) between $\tilde{X} | X_{\cR} = z$ and $Y$, taken over all uniformly random linear codes $C = [n, m]_q$, satisfies
\vspace{-0.1cm}
\begin{align}\label{eq:E_smooth_cond_bound_p}
\mathbb{E}\left[\mathbb{V}_{p}\left(p_{\tilde{X}|X_{\cR}=z},p_{Y}\right)\right] \leq 2^{\frac{p-1}{p}}\left((1+\epsilon)^p-1\right)^{\frac{1}{p}}.
\end{align}

That is, for a fixed $\mathcal{R} \in \binom{[n]}{r}$, a fixed $z \in \mathbb{F}_{q}^{r}$, and a key size~$m$ upholding~\eqref{eq:key_size}, there exists a linear code $C=[n,m]_{q}$ defined by a generator matrix $G \in \mathbb{F}_{q}^{m \times n}$ s.t.
\vspace{-0.1cm}
\begin{gather}\label{eq:smooth_cond_bound_p}
\mathbb{V}_{p}\left(p_{\tilde{X}|X_{\cR}=z},p_{Y}\right) \leq 2^{\frac{p-1}{p}}\left((1+\epsilon)^p-1\right)^{\frac{1}{p}}.
\end{gather}
Therefore, to ensure~\eqref{eq:smooth_cond_bound_p} for every~$\cR\in\binom{[n]}{r}$ and every~$z\in\bF_q^r$, we require
\vspace{-0.1cm}
\begin{small}
\begin{align}
    &m \geq \max_{\cR,z \in \mathbb{F}_{q}^{r}} \left( n + p + \log_{q}\left(\frac{1}{\epsilon}\right) - H_{p}(X|X_{\cR} = z) \right)\nonumber \\
    & = n + p + \log_{q}\left(\frac{1}{\epsilon}\right) - \min_{\cR,z \in \mathbb{F}_{q}^{r}}\left(H_{p}(X|X_{\cR} = z)\right) \nonumber \\
    & \quad \overset{(a)}{\geq} n + p + \log_{q}\left(\frac{1}{\epsilon}\right) -H_{p}(X) + \max_{\cR}H_{p}(X_{\cR}) \nonumber \\
    & \quad \overset{(b)}{=} n + p + \log_{q}\left(\frac{1}{\epsilon'}\right) -H_{p}(X), \label{eq:m_for_tilde_X}
\end{align}
\end{small}
\hspace{-0.15cm}where $(a)$ is shown in \cite[Appendix B]{ICC2025}, and~$(b)$ follows by denoting $\epsilon' = \epsilon q^{- \max_{\cR} H_{p}(X_{\cR})}$.

\off{according to \cite[Thm.~3.1, Lem.~3.6]{pathegama2024r}}
Since $m$ is lower bounded by~\eqref{eq:m_for_tilde_X}, the chosen random linear code $C$ ensures that the distribution of $\tilde{X}$ is nearly uniform over $\mathbb{F}_{q}^{n}$ as well. Thus, we have
\vspace{-0.1cm}
\begin{gather} \label{eq:smooth_bound_p}
\mathbb{V}_{p}\left(p_{\tilde{X}},p_{Y}\right) \leq 2^{\frac{p-1}{p}}\left((1+\epsilon')^p-1\right)^{\frac{1}{p}}.
\end{gather}

Finally, since $\mathbb{V}(\cdot,\cdot) \leq \mathbb{V}_{p}(\cdot,\cdot)$, we have that
\vspace{-0.1cm}
\begin{align}
\mathbb{V}\left(p_{\tilde{X}|X_{\mathcal{R}}=z},p_{Y}\right) &\leq \mathbb{V}_{p}\left(p_{\tilde{X}|X_{\mathcal{R}}=z},p_{Y}\right), \mbox{ and}\label{eq:cond_bound_p}\\ 
\mathbb{V}\left(p_{\tilde{X}},p_{Y}\right) &\leq \mathbb{V}_{p}\left(p_{\tilde{X}},p_{Y}\right) \label{eq:bound_p}.
\end{align}

Now, by applying~\eqref{eq:smooth_cond_bound_p},~\eqref{eq:smooth_bound_p},~\eqref{eq:cond_bound_p}, and~\eqref{eq:bound_p} over~\eqref{eq:k_s_ind}, we have that every~$\mathcal{R} \in \binom{[n]}{r}$ and every~$ z \in \mathbb{F}_{q}^{r}$ satisfy
\vspace{-0.1cm}
\begin{small}
\begin{align}
&\mathbb{V}\left(p_{\tilde{X}|X_{\mathcal{R}}=z},p_{\tilde{X}}\right) \nonumber  \\
&  \leq 2^{\frac{p-1}{p}}\left[\left((1+\epsilon)^p-1\right)^{\frac{1}{p}} + \left((1+\epsilon')^p-1\right)^{\frac{1}{p}}\right] \nonumber \\
&  \overset{(a)}{\leq} 2^{\frac{p-1}{p}}\left[2 \epsilon^{\frac{1}{p}} + 2 \epsilon'^{\frac{1}{p}}\right] = 2^{\frac{2p-1}{p}}\left(1 + q^{-\frac{\max_{\cR} H_{p}(X_{\cR})}{p}}\right)\epsilon^{\frac{1}{p}} \label{eq:v_bound},
\end{align}
\end{small}
\hspace{-0.15cm}where~$(a)$ follows since $\epsilon < 1$. The choice of $\epsilon$ affects both the size of the uniform key $K \in \mathbb{F}_{q}^{m}$ and the amount of information leakage due to the non-uniformity of $X$.
\off{According to \cite[Thm~3.1, Thm~3.6]{pathegama2024r}, for our choice of $m$ we obtain 
\vspace{-0.1cm}
\begin{gather}
    \mathbb{E}\left[\mathbb{V}\left(p_{\tilde{X}|X_{\mathcal{R}}=z},p_{\tilde{Y}|Y_{\mathcal{R}}=z}\right)\right] \leq 2^{\frac{p-1}{p}}\left((1+\epsilon)^p-1\right)^{\frac{1}{p}} \label{eq:expectation_V}
\end{gather}
for all~$z \in \mathbb{F}_{q}^{r}$, where the expectation is taken over all $[n,m]_{q}$ random linear codes.} 
Now, for $m$ that satisfies~\eqref{eq:m_for_tilde_X}, and for~$a > 1$, denote
\begin{small}
\begin{align*}
    E= \left\{C : \mathbb{V}\left(p_{\tilde{X}|X_{\mathcal{R}}=z},p_{Y}\right)\right. \left.\leq  a \cdot 2^{\frac{p-1}{p}}\left((1+\epsilon)^p-1\right)^{\frac{1}{p}}, \forall z,\cR\right\}.
\end{align*}
\end{small}
Further, notice that for $\mathcal{R}',z' = \argmax_{\mathcal{R},z}\max_{C\in E}\mathbb{V}\left(p_{\tilde{X}|X_{\mathcal{R}=z}},p_{Y}\right)$ 
we have that~$E$ can be written as
\begin{small}
\begin{align*}
    E= \left\{C : \mathbb{V}\left(p_{\tilde{X}|X_{\mathcal{R}'}=z'},p_{Y}\right)\right. \left.\leq  a \cdot 2^{\frac{p-1}{p}}\left((1+\epsilon)^p-1\right)^{\frac{1}{p}}\right\}.
\end{align*}
\end{small}

Thus, from~~\eqref{eq:E_smooth_cond_bound_p},~\eqref{eq:m_for_tilde_X},~\eqref{eq:cond_bound_p}, and Markov's inequality, we obtain $\Pr\left[E\right] \geq 1 - \frac{1}{a}$. That is, by choosing appropriate~$m$ and~$a$, with probability at least~$1-1/a$, a randomly chosen linear code $C=[n,m]_q$ satisfies
\vspace{-0.1cm}
\begin{align} \label{eq:final_leakage}
&\mathbb{V}\left(p_{\tilde{X}|X_{\mathcal{R}}=z},p_{\tilde{X}}\right) \leq a\cdot 2^{\frac{2p-1}{p}}\left(1 + q^{-\frac{\max_{\cR} H_{p}(X_{\cR})}{p}}\right)\epsilon^{\frac{1}{p}}
\end{align}
for all~$z \in \mathbb{F}_{q}^{r}$ and all~$\mathcal{R} \in \binom{[n]}{r}$.

\subsection{Leakage Proof of Theorem~\ref{theorem:main_theorem} Under Mutual Information}\label{sec:MI_theorem}
\off{We denote by $\delta = 2^{\frac{2p-1}{p}}(1 + q^{-\frac{\max_{r \in \mathcal{R}} H_{p}(X_r)}{p}})\epsilon^{\frac{1}{p}}$.} 
In this section, we demonstrate that the mutual information between $\tilde{X}$ and $X_{\mathcal{R}}$ is negligible for all~$\cR$, thereby ensuring that the proposed protocol satisfies the ICC conditions (Definition~\ref{definition:ICC}). This is established using the equivalence between $p$-total variation and $p$-KL Divergence, as shown in \cite[Prop.~2.1]{pathegama2024r}.

First, we denote by $\delta = a \cdot 2^{\frac{2p-1}{p}}\left(1 + q^{-\frac{\max_{\cR} H_{p}(X_{\cR})}{p}}\right)\epsilon^{\frac{1}{p}}$.
Directly applying \cite[Prop~2.1]{pathegama2024r} (see \cite[eq.~(12)]{ICC2025}) on~\eqref{eq:final_leakage} gives us that every~$\mathcal{R} \in \binom{[n]}{r}$ and every~$z \in \mathbb{F}_{q}^{r}$ satisfy
\vspace{-0.1cm}
\begin{gather}
\mathbb{D}_{p}\left(p_{\tilde{X}|X_{\mathcal{R}}=z}||p_{\tilde{X}}\right) \leq \frac{p}{p-1} \log_{q}(1 + \delta). \label{eq:D_cond_bound}
\end{gather}

To bound $I(\tilde{X};X_{\cR})$, recall the equivalence between the mutual information and KL-Divergence: $I(\tilde{X};X_{\mathcal{R}}) = \mathbb{D}\left(p_{\tilde{X},X_{\mathcal{R}}}||p_{\tilde{X}}p_{X_{\mathcal{R}}}\right)$. 
Since $\mathbb{D}\left(p_{\tilde{X},X_{\mathcal{R}}}||p_{\tilde{X}}p_{X_{\mathcal{R}}}\right) \leq \mathbb{D}_{p}\left(p_{\tilde{X},X_{\mathcal{R}}}||p_{\tilde{X}}p_{X_{\mathcal{R}}}\right)$, we focus on bounding $\mathbb{D}_{p}\left(p_{\tilde{X},X_{\mathcal{R}}}||p_{\tilde{X}}p_{X_{\mathcal{R}}}\right)$ as follows
\vspace{-0.1cm}
\begin{small}
\begin{align*}
&\mathbb{D}_{p}\left(p_{\tilde{X},X_{\cR}}||p_{\tilde{X}}p_{X_{\mathcal{R}}}\right) \nonumber \\
& \quad \overset{(a)}{=} \frac{1}{p-1} \log_{q}\left( \sum_{\tilde{x},z}p^{p}_{\tilde{X},X_\cR}(\tilde{x},z) p^{-(p-1)}_{\tilde{X}}(\tilde{x})p^{-(p-1)}_{X_\cR}(z)\right) \nonumber \\
& \quad \overset{(b)}{=} \frac{1}{p-1} \log_{q}\left( \sum_{z}p_{X_\cR}(z)\sum_{\tilde{x}}p^{p}_{\tilde{X}|X_\cR}(\tilde{x}|z) p^{-(p-1)}_{\tilde{X}}(\tilde{x})\right) \nonumber \\
& \quad \overset{(c)}{=} \frac{1}{p-1} \log_{q}\left( \sum_{z}p_{X_\cR}(z)q^{(p-1)\mathbb{D}_{p}\left(p_{\tilde{X}|X_\cR=z}||p_{\tilde{X}}\right)}\right) \nonumber \\
& \quad \overset{(d)}{\leq} \frac{1}{p-1} \log_{q}\left( \sum_{z}p_{X_\cR}(z)q^{p\log_{q}(1+\delta)}\right) \nonumber \\
& \quad = \frac{p}{p-1}\log_{q}(1 + \delta),
\end{align*}
\end{small}
\hspace{-0.15cm}where~$(a)$, $(b)$, and~$(c)$ follow from the definition of $\mathbb{D}_{p}(\cdot||\cdot)$, and~$(d)$ follows from \eqref{eq:D_cond_bound}. Thus, using the equivalence between mutual information and KL-Divergence, as well as the relation $\mathbb{D}(\cdot||\cdot) \leq \mathbb{D}_{p}(\cdot||\cdot)$, we obtain
    $I(\tilde{X};X_{\mathcal{R}}) \leq \frac{p}{p - 1} \log_{q}(1 + \delta)$
for all~$\cR$, and hence ICC in Theorem~\ref{theorem:main_theorem} is satisfied\off{ with~$\epsilon=\frac{p}{p-1}\log_q(1+\delta)$}.

\newpage
\bibliographystyle{IEEEtran}
\bibliography{refs}

\begin{thebibliography}{10}
\providecommand{\url}[1]{#1}
\csname url@samestyle\endcsname
\providecommand{\newblock}{\relax}
\providecommand{\bibinfo}[2]{#2}
\providecommand{\BIBentrySTDinterwordspacing}{\spaceskip=0pt\relax}
\providecommand{\BIBentryALTinterwordstretchfactor}{4}
\providecommand{\BIBentryALTinterwordspacing}{\spaceskip=\fontdimen2\font plus
\BIBentryALTinterwordstretchfactor\fontdimen3\font minus \fontdimen4\font\relax}
\providecommand{\BIBforeignlanguage}[2]{{%
\expandafter\ifx\csname l@#1\endcsname\relax
\typeout{** WARNING: IEEEtran.bst: No hyphenation pattern has been}%
\typeout{** loaded for the language `#1'. Using the pattern for}%
\typeout{** the default language instead.}%
\else
\language=\csname l@#1\endcsname
\fi
#2}}
\providecommand{\BIBdecl}{\relax}
\BIBdecl

\bibitem{cuff2016differential}
P.~Cuff and L.~Yu, ``Differential privacy as a mutual information constraint,'' in \emph{Proceedings of the 2016 ACM SIGSAC Conference on Computer and Communications Security}, 2016, pp. 43--54.

\bibitem{mironov2009computational}
I.~Mironov, O.~Pandey, O.~Reingold, and S.~Vadhan, ``Computational differential privacy,'' in \emph{Annual International Cryptology Conference}.\hskip 1em plus 0.5em minus 0.4em\relax Springer, 2009, pp. 126--142.

\bibitem{huang2019dp}
Z.~Huang, R.~Hu, Y.~Guo, E.~Chan-Tin, and Y.~Gong, ``{DP-ADMM: ADMM-based distributed learning with differential privacy},'' \emph{IEEE Transactions on Information Forensics and Security}, vol.~15, pp. 1002--1012, 2019.

\bibitem{li2009tradeoff}
T.~Li and N.~Li, ``On the tradeoff between privacy and utility in data publishing,'' in \emph{Proceedings of the 15th ACM SIGKDD international conference on Knowledge discovery and data mining}, 2009, pp. 517--526.

\bibitem{wang2017privacy}
Y.~Wang, Y.~O. Basciftci, and P.~Ishwar, ``Privacy-utility tradeoffs under constrained data release mechanisms,'' \emph{arXiv preprint arXiv:1710.09295}, 2017.

\bibitem{calmon2015fundamental}
F.~P. Calmon, A.~Makhdoumi, and M.~M{\'e}dard, ``Fundamental limits of perfect privacy,'' in \emph{2015 IEEE International Symposium on Information Theory (ISIT)}.\hskip 1em plus 0.5em minus 0.4em\relax IEEE, 2015, pp. 1796--1800.

\bibitem{rassouli2017perfect}
B.~Rassouli and D.~G{\"u}nd{\"u}z, ``On perfect privacy and maximal correlation,'' \emph{arXiv preprint arXiv:1712.08500}, vol.~2, no.~3, 2017.

\bibitem{rassouli2021perfect}
------, ``On perfect privacy,'' \emph{IEEE Journal on Selected Areas in Information Theory}, vol.~2, no.~1, pp. 177--191, 2021.

\bibitem{mulligan2021confidential}
D.~P. Mulligan, G.~Petri, N.~Spinale, G.~Stockwell, and H.~J. Vincent, ``Confidential computing—a brave new world,'' in \emph{2021 international symposium on secure and private execution environment design (SEED)}.\hskip 1em plus 0.5em minus 0.4em\relax IEEE, 2021, pp. 132--138.

\bibitem{hunt2021confidential}
G.~D. Hunt, R.~Pai, M.~V. Le, H.~Jamjoom, S.~Bhattiprolu, R.~Boivie, L.~Dufour, B.~Frey, M.~Kapur, K.~A. Goldman \emph{et~al.}, ``Confidential computing for openpower,'' in \emph{Proceedings of the Sixteenth European Conference on Computer Systems}, 2021, pp. 294--310.

\bibitem{feng2024survey}
D.~Feng, Y.~Qin, W.~Feng, W.~Li, K.~Shang, and H.~Ma, ``Survey of research on confidential computing,'' \emph{IET Communications}, vol.~18, no.~9, pp. 535--556, 2024.

\bibitem{raviv2022perfect}
N.~Raviv and Z.~Goldfeld, ``Perfect subset privacy for data sharing and learning,'' in \emph{2022 IEEE International Symposium on Information Theory (ISIT)}.\hskip 1em plus 0.5em minus 0.4em\relax IEEE, 2022, pp. 1850--1855.

\bibitem{lee2017speeding}
K.~Lee, M.~Lam, R.~Pedarsani, D.~Papailiopoulos, and K.~Ramchandran, ``Speeding up distributed machine learning using codes,'' \emph{IEEE Transactions on Information Theory}, vol.~64, no.~3, pp. 1514--1529, 2017.

\bibitem{yu2019lagrange}
Q.~Yu, S.~Li, N.~Raviv, S.~M.~M. Kalan, M.~Soltanolkotabi, and S.~A. Avestimehr, ``Lagrange coded computing: Optimal design for resiliency, security, and privacy,'' in \emph{The 22nd International Conference on Artificial Intelligence and Statistics}.\hskip 1em plus 0.5em minus 0.4em\relax PMLR, 2019, pp. 1215--1225.

\bibitem{d2020gasp}
R.~G. D’Oliveira, S.~El~Rouayheb, and D.~Karpuk, ``{GASP} codes for secure distributed matrix multiplication,'' \emph{IEEE Transactions on Information Theory}, vol.~66, no.~7, pp. 4038--4050, 2020.

\bibitem{wang2022breaking}
C.~Wang and N.~Raviv, ``Breaking blockchain’s communication barrier with coded computation,'' \emph{IEEE Journal on Selected Areas in Information Theory}, vol.~3, no.~2, pp. 405--421, 2022.

\bibitem{deng2024perfect}
Z.~Deng, V.~Ramkumar, and N.~Raviv, ``Perfect subset privacy in polynomial computation,'' in \emph{2024 IEEE International Symposium on Information Theory (ISIT)}.\hskip 1em plus 0.5em minus 0.4em\relax IEEE, 2024, pp. 933--938.

\bibitem{10619109}
S.~Tarnopolsky and A.~Cohen, ``Coding-based hybrid post-quantum cryptosystem for non-uniform information,'' in \emph{2024 IEEE International Symposium on Information Theory (ISIT)}, 2024, pp. 1830--1835.

\bibitem{bellare1998relations}
M.~Bellare, A.~Desai, D.~Pointcheval, and P.~Rogaway, ``Relations among notions of security for public-key encryption schemes,'' in \emph{Advances in Cryptology—CRYPTO'98: 18th Annual International Cryptology Conference Santa Barbara, California, USA August 23--27, 1998 Proceedings 18}.\hskip 1em plus 0.5em minus 0.4em\relax Springer, 1998, pp. 26--45.

\bibitem{enwiki:indistinguishability}
\BIBentryALTinterwordspacing
{Wikipedia contributors}, ``{Ciphertext indistinguishability} --- {Wikipedia}{,} the free encyclopedia,'' 2024, [Online; accessed September 2024]. [Online]. Available: \url{https://en.wikipedia.org/wiki/Ciphertext_indistinguishability}
\BIBentrySTDinterwordspacing

\bibitem{enwiki:Negligible}
\BIBentryALTinterwordspacing
------, ``{Negligible function} --- {Wikipedia}{,} the free encyclopedia,'' 2024, [Online; accessed December 2024]. [Online]. Available: \url{https://en.wikipedia.org/wiki/Negligible_function}
\BIBentrySTDinterwordspacing

\bibitem{pathegama2024r}
M.~Pathegama and A.~Barg, ``R$\backslash$'enyi divergence guarantees for hashing with linear codes,'' \emph{arXiv preprint arXiv:2405.04406}, 2024.

\bibitem{shannon1949communication}
C.~E. Shannon, ``Communication theory of secrecy systems,'' \emph{Bell system technical journal}, vol.~28, no.~4, pp. 656--715, 1949.

\bibitem{bloch2011physical}
M.~Bloch and J.~Barros, \emph{Physical-Layer Security: From Information Theory to Security Engineering}.\hskip 1em plus 0.5em minus 0.4em\relax Cambridge University Press, 2011.

\bibitem{ozarow1985wire}
L.~H. Ozarow and A.~D. Wyner, ``Wire-tap channel ii,'' in \emph{Advances in Cryptology}.\hskip 1em plus 0.5em minus 0.4em\relax Springer, 1985, pp. 33--50.

\bibitem{carleial1977note}
A.~Carleial and M.~Hellman, ``A note on wyner's wiretap channel (corresp.),'' \emph{IEEE Transactions on Information Theory}, vol.~23, no.~3, pp. 387--390, 1977.

\bibitem{bhattad2005weakly}
K.~Bhattad, K.~R. Narayanan \emph{et~al.}, ``Weakly secure network coding,'' \emph{NetCod, Apr}, vol. 104, pp. 8--20, 2005.

\bibitem{cohen2018secure}
A.~Cohen, A.~Cohen, M.~Medard, and O.~Gurewitz, ``Secure multi-source multicast,'' \emph{IEEE Transactions on Communications}, vol.~67, no.~1, pp. 708--723, 2018.

\bibitem{matsumoto2017universal}
R.~Matsumoto and M.~Hayashi, ``Universal secure multiplex network coding with dependent and non-uniform messages,'' \emph{IEEE Transactions on Inf. Theory}, vol.~63, no.~6, pp. 3773--3782, 2017.

\bibitem{sevilla2022compute}
J.~Sevilla, L.~Heim, A.~Ho, T.~Besiroglu, M.~Hobbhahn, and P.~Villalobos, ``Compute trends across three eras of machine learning,'' in \emph{2022 International Joint Conference on Neural Networks (IJCNN)}.\hskip 1em plus 0.5em minus 0.4em\relax IEEE, 2022, pp. 1--8.

\bibitem{egger2022efficient}
M.~Egger, R.~Bitar, A.~Wachter-Zeh, and D.~G{\"u}nd{\"u}z, ``Efficient distributed machine learning via combinatorial multi-armed bandits,'' in \emph{2022 IEEE International Symposium on Information Theory (ISIT)}.\hskip 1em plus 0.5em minus 0.4em\relax IEEE, 2022, pp. 1653--1658.

\bibitem{xhemrishi2022computational}
M.~Xhemrishi, A.~G. i~Amat, E.~Rosnes, and A.~Wachter-Zeh, ``Computational code-based privacy in coded federated learning,'' in \emph{2022 IEEE International Symposium on Information Theory (ISIT)}.\hskip 1em plus 0.5em minus 0.4em\relax IEEE, 2022, pp. 2034--2039.

\bibitem{dwork2006differential}
C.~Dwork, ``Differential privacy,'' in \emph{International colloquium on automata, languages, and programming}.\hskip 1em plus 0.5em minus 0.4em\relax Springer, 2006, pp. 1--12.

\bibitem{rassouli2019data}
B.~Rassouli, F.~E. Rosas, and D.~G{\"u}nd{\"u}z, ``Data disclosure under perfect sample privacy,'' \emph{IEEE Transactions on Information Forensics and Security}, vol.~15, pp. 2012--2025, 2019.

\bibitem{micciancio2007worst}
D.~Micciancio and O.~Regev, ``Worst-case to average-case reductions based on gaussian measures,'' \emph{SIAM Journal on Computing}, vol.~37, no.~1, pp. 267--302, 2007.

\bibitem{debris2023smoothing}
T.~Debris-Alazard, L.~Ducas, N.~Resch, and J.-P. Tillich, ``Smoothing codes and lattices: Systematic study and new bounds,'' \emph{IEEE Transactions on Information Theory}, vol.~69, no.~9, pp. 6006--6027, 2023.

\bibitem{ICC2025}
S.~Tarnopolsky, Z.~Deng, V.~Ramkumar, N.~Raviv, and A.~Cohen, ``Individual confidential computing of polynomials over non-uniform information,'' \url{https://drive.google.com/drive/folders/1pw1ln2emJ5leEHb1_WKcv9IdRtb6-fOj?usp=sharing}, 2025.

\bibitem{NegligbleCost}
R.~A. Chou, B.~N. Vellambi, M.~R. Bloch, and J.~Kliewer, ``Coding schemes for achieving strong secrecy at negligible cost,'' \emph{IEEE Transactions on Information Theory}, vol.~63, no.~3, pp. 1858--1873, 2017.

\bibitem{1053968}
S.~Kullback, ``A lower bound for discrimination information in terms of variation (corresp.),'' \emph{IEEE Trans. on Inf. Theory}, vol.~13, no.~1, pp. 126--127, 1967.

\end{thebibliography}
\newpage
\appendices
\section{Smoothing} \label{appendix:smoothing}
In this paper, we utilize smoothing of distributions \cite{micciancio2007worst,debris2023smoothing} as a method to achieve perfect subset privacy for non-uniform information. Smoothing of distributions is used as a technique to transform some non-uniform random variable $X \in \mathbb{F}_{q}^{n}$ into an almost uniform one.

We now introduce several useful information metrics that are utilized throughout this paper to measure distances between probability distributions. Let $p_1$ and $p_2$ be two distinct distributions over $\mathbb{F}_{q}^{n}$. First, we consider the variational distance and the $p$-variational distance for $p \in \mathbb{N}$, $p \geq 2$, defined between $p_1$ and $p_2$ as follows, and defined as follows
\vspace{-0.1cm}
\begin{gather*}
    \mathbb{V}(p_1,p_2) = \sum_{x \in \mathbb{F}_{q}^{n}}|p_{1}(x) - p_{2}(x)|, 
\end{gather*}
\vspace{-0.15cm}
\begin{gather*}
    \mathbb{V}_{p}(p_1,p_2) = q^{n}\left(\frac{1}{q^n}\sum_{x \in \mathbb{F}_{q}^{n}}|p_{1}(x) - p_{2}(x)|^{p}\right)^{\frac{1}{p}},    
\end{gather*}
where $\mathbb{V}(\cdot,\cdot)$ is the variational distance and $\mathbb{V}_{p}(\cdot,\cdot)$ is the $p$-variational distance. Since convergence in $\ell_p$ is stronger than convergence in $\ell_1$, we get $\mathbb{V}(\cdot,\cdot) \leq \mathbb{V}_{p}(\cdot,\cdot)$.

Another information metric considered in this paper is the KL-Divergence, along with the $p$-KL-Divergence for $p \in \mathbb{N}$, $p \geq 2$, which are defined as
\vspace{-0.1cm}
\begin{gather*}
    \mathbb{D}(p_1||p_2) = \sum_{x \in \mathbb{F}_{q}^{n}} p_{1}(x) \log_{q} \frac{p_{1}(x)}{p_{2}(x)},    
\end{gather*}
\vspace{-0.25cm}
\begin{gather*}
    \mathbb{D}_{p}(p_1||p_2) = \frac{1}{p-1}\log_{q}\left(\sum_{x \in \mathbb{F}_{q}^{n}}p^{p}_{1}(x) p^{-(p-1)}_{2}(x)\right),    
\end{gather*}
where $\mathbb{D}(\cdot||\cdot)$ is the KL-Divergence and $\mathbb{D}_{p}(\cdot||\cdot)$ is the $p$-KL-Divergence. Extending this notation to entropy, we have the following definition for $p$-entropy of the random variable~$X$
\vspace{-0.1cm}
\begin{gather*}
    H_{p}(X) = \frac{1}{p-1}\log_{q}\left(\sum_{x \in \mathbb{F}_{q}^{n}}p^{p}(x)\right).   
\end{gather*}

We note that $p$-entropy is a decreasing function of $p$, whereas $p$-KL-Divergence is an increasing function of $p$.

Both $p$-variational distance and $p$-KL Divergence are  metrics often used to quantify information leakage~\cite{bloch2011physical,NegligbleCost}. For $p = 1$, the relation between the two metrics is defined by the Pinsker inequality \cite{1053968} as
    $\mathbb{V}(p_1,p_2) \leq \sqrt{\frac{1}{2}\mathbb{D}(p_1||p_2)}$. 

That is, if the KL-Divergence between two distributions is bounded, then the variational distance between them can also be bounded, but the other direction is not necessarily true. 
However, in \cite[Prop. 2.1]{pathegama2024r} a two-way relation between the two metrics is introduced for $p \geq 2$. More specifically, it is shown that
\begin{align}\label{equation:Sasha'sIneq}
    \mathbb{V}_{p}(p_1,p_2) \leq \delta \Rightarrow \mathbb{D}_{p}(p_1||p_2) \leq \frac{p}{p-1}\log_{q}(1+\delta).
\end{align}

In this paper, we consider smoothing of distributions using random linear codes (Definition~\ref{def:smoothing}).
\off{\begin{definition}\label{def:smoothing}
    (Smoothing using Linear Codes) Let $X \in \mathbb{F}_{q}^{n}$ be some non-uniform random variable with distribution $X \sim P$. Let $C = [n,m]_{q}$ a uniformly chosen linear code. Let $\mathcal{C} \in \mathbb{F}_{q}^{n}$ be a randomly chosen codeword from~$C$, and $\tilde{X} = X + \mathcal{C}$. Denote the distribution $\tilde{X} \sim \tilde{P}$. We say the code $C$ is smoothing the distribution $X \sim P$ if
        $\mathbb{V}_{p}\left(\tilde{P},P_{U_n}\right) \leq \epsilon$
    for some negligible $\epsilon > 0$, where $P_{U_n}$ is the uniform distribution over $\mathbb{F}_{q}^{n}$.
\end{definition}}
Thus, the encoding $\tilde{X} = X + \mathcal{C}$, where $\mathcal{C}$ is chosen uniformly from $C$, generates a random variable $\tilde{X}$ that is nearly uniformly distributed over $\mathbb{F}_{q}^{n}$.

In \cite{pathegama2024r}, Panthegama et. al. provide the conditions and encoding for which a random linear code can perform smoothing on a non-uniform variable~\cite[Thm.~3.1]{pathegama2024r}. 
The theorem states that there exists a random linear code $C = [n,m]_q$ with a generator matrix $G \in \mathbb{F}_{q}^{m \times n}$ for which applying the linear transformation $\tilde{X} = X + KG$ on the non-uniform random variable $X \in \mathbb{F}_q^n$ using a random key $K \sim \text{Unif}(\mathbb{F}_q^m)$ results in a nearly uniformly distributed random variable $\tilde{X}  \in \mathbb{F}_q^n$. Specifically, the theorem asserts that for integer $p \geq 2$ and negligible $\epsilon > 0$, taking $m$ larger than a function of $n, p, \log_q(1/\epsilon)$, and $H_p(X)$ ensures the existence of a linear code $C = [n,m]_q$ s.t. the $p$-variational distance between the distributions $p_{X}$ and $p_{\tilde{X}}$ is 
\vspace{-0.1cm}
\begin{gather*}
    \mathbb{V}_{p}(p_{\tilde{X}},p_{U_n}) \leq 2^{\frac{p-1}{p}}((1+\epsilon)^{p}-1)^{\frac{1}{p}}.
\end{gather*}

We apply this result within the context of ICC (Definition~\ref{definition:ICC}). ICC quantifies information leakage through mutual information $I(\tilde{X};X_{\mathcal{R}})$, and requires this term to be bounded by some negligible value for all $\cR \in \binom{[n]}{r}$, where $\binom{[n]}{r}$ is the collection of all subsets of size~$1 \leq r < n$\off{ of~$[n]$}. 
We recall that mutual information is equivalent to KL-Divergence, i.e., $I(\tilde{X};X_{\mathcal{R}}) = \mathbb{D}(p_{\tilde{X},X_{\mathcal{R}}}||p_{\tilde{X}}p_{X_{\cR}})$. 
Additionally, we recall that for the mutual information between two random variables (or the KL-divergence between two distributions) to approach zero, the random variables need to be almost independent of each other. In terms of variational distance, this requirement is equivalent to requiring that the variational distance between the distributions $p_{\tilde{X}}$ and $p_{\tilde{X}|X_{\mathcal{R}}}$ be negligible. In Sec.~\ref{sec:VD_theorem} and~\ref{sec:MI_theorem}, smoothing  and \cite[Thm.~3.1]{pathegama2024r} are utilized to make the required distributions nearly uniform, i.e., make $\mathbb{V}(p_{\tilde{X}},p_{U_n})$ and $\mathbb{V}(p_{X_{\mathcal{R}}},p_{U_n})$ negligible, where $p_{U_n}$ is the uniform distribution over $F_{q}^{n}$. Hence, the distributions $p_{\tilde{X}}$ and $p_{X_{\mathcal{R}}}$ are close to each other as well, which in turn ensures negligible leakage and satisfaction of ICC conditions. In Sec.~\ref{sec:MI_theorem} this result is expanded to the mutual information $I(\tilde{X},X_{\mathcal{R}})$ as well.
\section{Inequality Proof}\label{appendix:proof_H_ineq}
In this section, we prove inequality~$(a)$ of~\eqref{eq:m_for_tilde_X} in Sec.~\ref{sec:VD_theorem}. We denote by $v = n - r$ and $\mathcal{V} = \mathcal{R}^{C}$, as well as $p(x)\triangleq p_{X}(X=x)$, $p(z)\triangleq p_{X_\mathcal{R}}(X_\mathcal{R} = z)$ and $p(y|z)\triangleq p_{X_\mathcal{V}|X_\mathcal{R}=z}(X_\mathcal{V} = y|X_\mathcal{R} = z)$, 
where $x \in \mathbb{F}_{q}^{n}$, $z \in \mathbb{F}_{q}^{r}$, and $y \in \mathbb{F}_{q}^{v}$. 

First, we consider the following bound for a fixed $\mathcal{R} \in \binom{[n]}{r}$.
\begin{align}
    &H_{p}(X) - H_{p}(X_{\mathcal{R}}) - \min_{z} \left(H_{p}(X|X_{\mathcal{R}}=z)\right) \nonumber \\
    & \quad \overset{(a)}{=} \frac{1}{1-p}\left(\log_{q}\left(\sum_{x \in \mathbb{F}_{q}^{n}}p^{p}(x)\right) - \log_{q}\left(\sum_{z \in \mathbb{F}_{q}^{r}}p^{p}(z)\right)\right) \nonumber \\
    & \quad\quad\quad\quad\quad\quad\quad\quad\quad\quad\quad\quad\quad\quad -\min_{z} \left(H_{p}(X|X_{\mathcal{R}}=z)\right) \nonumber \\
    & \quad = \frac{1}{p-1}\log_{q}\left(\frac{\sum_{z \in \mathbb{F}_{q}^{r}}p^{p}(z)}{\sum_{x \in \mathbb{F}_{q}^{n}}p^{p}(x)}\right) -\min_{z} \left(H_{p}(X|X_{\mathcal{R}}=z)\right) \nonumber \\
    & \quad = \frac{1}{p-1}\log_{q}\left(\frac{\sum_{z \in \mathbb{F}_{q}^{r}}p^{p}(z)}{\sum_{z \in \mathbb{F}_{q}^{r}}p^{p}(z)\sum_{y \in \mathbb{F}_{q}^{v}}p^{p}(y|z)}\right) \nonumber \\
    & \quad\quad\quad\quad\quad\quad\quad\quad\quad\quad\quad -\min_{z} \left(H_{p}(X|X_{\mathcal{R}}=z)\right) \nonumber \\
    & \quad \overset{(b)}{=} \frac{1}{p-1}\log_{q}\left(\frac{\sum_{z \in \mathbb{F}_{q}^{r}}p^{p}(z)}{\sum_{z \in \mathbb{F}_{q}^{r}}p^{p}(z)\sum_{y \in \mathbb{F}_{q}^{v}}p^{p}(y|z)}\right) \nonumber \\
    & \quad\quad\quad -\min_{z} \left(\frac{1}{p-1} \log_{q}\left(\frac{1}{\sum_{y \in \mathbb{F}_{q}^{v}}p^{p}(y|z)}\right)\right) \nonumber \\
    & \quad \overset{(c)}{=} \frac{1}{p-1}\log_{q}\left(\frac{\sum_{z \in \mathbb{F}_{q}^{r}}p^{p}(z)}{\sum_{z \in \mathbb{F}_{q}^{r}}p^{p}(z)\sum_{y \in \mathbb{F}_{q}^{v}}p^{p}(y|z)}\right) \nonumber \\
    & \quad\quad\quad -\frac{1}{p-1} \log_{q}\left(\frac{1}{\sum_{y \in \mathbb{F}_{q}^{v}}p^{p}(y|z')}\right) \nonumber \\
    & = \frac{1}{p-1}\log_{q}\left(\frac{\sum_{z \in \mathbb{F}_{q}^{r}}p^{p}(z)\sum_{y \in \mathbb{F}_{q}^{v}}p^{p}(y|z')}{\sum_{z \in \mathbb{F}_{q}^{r}}p^{p}(z)\sum_{y \in \mathbb{F}_{q}^{v}}p^{p}(y|z)}\right) \overset{(d)}{\geq} 0, \nonumber
\end{align}
where~$(a)$ and~$(b)$ follow from the definition of~$H_{p}(\cdot)$ (Sec.~\ref{sec:smoothing}), and~$(c)$ follows by setting $z'$ as the minimizer of $\min_{z \in \mathbb{F}_{q}^{r}}\left(H_{p}(X|X_{\mathcal{R}}=z)\right)$ for a fixed $\mathcal{R} \in \binom{[n]}{r}$. Since $z'$ is defined as a minimizer, and since $\log_{q}(\cdot)$ is a monotonically increasing function, it follows that $z'$ maximizes the expression $\sum_{y \in \mathbb{F}_{q}^{v}}p^{p}(y|z)$ over all $z \in \mathbb{F}_{q}^{r}$. Thus we have inequality~$(d)$, which is a result of the logarithmic argument being greater than one. 
That is, we have shown that for a fixed $\mathcal{R} \in \binom{[n]}{r}$,
\vspace{-0.1cm}
\begin{gather} \label{eq:first_min}
    H_{p}(X) - H_{p}(X_{\mathcal{R}}) \geq \min_{z}\left(H_{p}(X|X_{\mathcal{R}}=z)\right).
\end{gather}
Now, we minimize both sides over all $\mathcal{R} \in \binom{[n]}{r}$. On the left side we obtain
\vspace{-0.1cm}
\begin{gather}\label{eq:max_H_cond}
     \min_{\mathcal{R} \in \binom{[n]}{r}}\left(H_{p}(X) - H_{p}(X_{\mathcal{R}})\right) = H_{p}(X) - \max_{\mathcal{R} \in \binom{[n]}{r}}\left(H_{p}(X_{\mathcal{R}})\right).
\end{gather}
Finally, using~\eqref{eq:first_min} and~\eqref{eq:max_H_cond}, we have that
\vspace{-0.1cm}
\begin{align} 
    &H_{p}(X) - \max_{\mathcal{R} \in \binom{[n]}{r}}\left(H_{p}(X_{\mathcal{R}})\right) \nonumber \\ 
     & \quad\quad\quad\quad\quad\quad\quad \geq \min_{\cR \in \binom{[n]}{r},z \in \mathbb{F}_{q}^{r}}\left(H_{p}(X|X_{\cR} = z)\right).
\end{align}

\end{document}